\PassOptionsToPackage{table}{xcolor}
\documentclass[aps,prl,twocolumn,amsmath,superscriptaddress,preprintnumbers,amssymb,floatfix,footinbib,longbibliography]{revtex4-2}

\makeatletter
 
\renewcommand{\p@subsection}{}  
\makeatother
\usepackage{float}
\usepackage{graphicx}
\usepackage{amsmath,amssymb}
\usepackage{tikz}
\usepackage[compat=1.1.0]{tikz-feynman}
\usepackage{comment}

\usepackage{siunitx}

\usepackage{placeins}

\usepackage[colorlinks=true,linkcolor=blue,citecolor=blue,urlcolor=blue]{hyperref}

\usepackage[nameinlink,noabbrev]{cleveref}
\providecommand{\qty}[2]{\SI{#1}{#2}}
\providecommand{\qtyrange}[3]{\SIrange{#1}{#2}{#3}}
\usepackage{booktabs}
\usepackage[table]{xcolor}
\usepackage{makecell}
\usepackage{placeins}
\usepackage{cleveref}
\crefname{figure}{fig.}{figs.}
\Crefname{figure}{Fig.}{Figs.}

\crefname{equation}{eq.}{eqs.}
\Crefname{equation}{Eq.}{Eqs.}
\crefname{table}{tab.}{tabs.}
\Crefname{table}{Tab.}{Tabs.}

\tikzfeynmanset{ fermion/.style={
   decoration={
     markings,
     mark=at position 0.5 with {\arrow[xshift=1mm]{Latex[width=1.5mm,length=2mm]}}
   },
   postaction=decorate
}}
\tikzfeynmanset{ anti fermion/.style={
   decoration={
     markings,
     mark=at position 0.5 with {\arrowreversed[xshift=-1mm]{Latex[width=1.5mm,length=2mm]}}
   },
   postaction=decorate
}}

\tikzfeynmanset{warn luatex=false}
\begin{document}

\title{Cornering MeV-GeV Axions and Dark Photons with LDMX}

\author{Sarah Gaiser}
\affiliation{SLAC National Accelerator Laboratory, Menlo Park, CA 94025, USA}
\affiliation{Department of Physics, Stanford University, Stanford, CA 94305, USA}
\author{Alessandro Russo}
\thanks{arusso00@stanford.edu}
\affiliation{SLAC National Accelerator Laboratory, Menlo Park, CA 94025, USA}
\affiliation{Leinweber Institute for Theoretical Physics, Department of Physics, Stanford University, Stanford, CA 94305, USA}
\affiliation{Department of Physics, Stanford University, Stanford, CA 94305, USA}
\author{Philip Schuster}
\affiliation{SLAC National Accelerator Laboratory, Menlo Park, CA 94025, USA}

\begin{abstract}
Axion-like particles (ALPs), the QCD axion, and dark photons in the MeV-GeV mass range are motivated by various dark matter models and the strong CP problem, and are ubiquitous in extensions of the Standard Model. A long-standing blind spot for experimental searches is the sub-\qty{100}{MeV} mass range, where the particle lifetime is too long to be constrained by prompt-decay collider searches yet too short to be reached by beam-dump experiments. We investigate and estimate the sensitivity of the Light Dark Matter eXperiment (LDMX) to such axions and dark photons, motivated by the clean environment in which these particles can be produced and by the near-target tracking capabilities of LDMX. With reasonable charged track and momentum reconstruction capabilities, we find that LDMX could close much of this low-mass blind spot for axions and dark photons.
\end{abstract}

\maketitle

\section{Introduction}

The QCD axion and axion-like particles (ALPs) arise as pseudo–Nambu–Goldstone bosons associated with the spontaneous breaking of a global symmetry, while dark photons correspond to an additional Abelian gauge boson that kinetically mixes with the Standard Model photon. For many years, such particles in the MeV-GeV mass range have received considerable experimental and theoretical attention (see \cite{Gori:2022vri, OHare:2024nmr, Blinov:2022tfy, 10.21468/SciPostPhysRev.1, Batell:2022dpx, adams2023axiondarkmatter} for a recent review). However, fully constraining the sub-GeV mass range remains highly challenging, and many existing searches lack sensitivity across broad regions of parameter space, particularly in the sub-\qty{100}{MeV} regime~\cite{Berlin:2018bsc, Belle-II:2020jti, Jaeckel:2015jla}. 

In this work, we focus on the interactions of axions and dark photons with Standard Model fermions, specifically electrons and muons. We investigate the production and detection of axions and dark photons near the target of the Light Dark Matter eXperiment (LDMX) \cite{akesson2025ldmxlightdark}. We find that for reasonable tracking performance (similar to the performance of the Heavy Photon Search (HPS), see \cite{PhysRevD.108.012015,PhysRevD.98.091101,2k3q-9mhj}), LDMX's unique near-target tracking capability could close much of the unexplored parameter space below \qty{100}{MeV}.

In the following sections (Sec.~\ref{sec_axion} and Sec.~\ref{sec_DP}), we outline representative axion and dark photon models, followed by a short discussion of the LDMX setup in Sec.~\ref{sec_ldmx}. Section~\ref{sec_CALC} introduces 
the relevant signal and background processes, including models for their impact parameters and invariant mass distributions, before we
discuss the experimental geometry and the selection cuts in Sec. \ref{sec_geometry}.
The definition and optimization of the significance are presented in Sec.~\ref{sec_sig}.
Finally, our results are shown in Sec.~\ref{sec_RES}.

\subsection{Axion Models} \label{sec_axion}

Axions and ALPs are pseudoscalar states which, in most well-motivated extensions of the Standard Model~\cite{Blinov:2022tfy, Choi:2020rgn, Graham:2015ouw, Arza:2026rsl}, interact with Standard Model fields through higher-dimensional operators.
At low energies, their interactions with fermions can be described by the effective Lagrangian:

\begin{equation}
\label{eq:axion_lagrange}
\mathcal{L} \;\supset\; \sum_{f} \frac{m_f}{\Lambda_{f}}\, a\, \bar{f} \gamma^5 f \, ,
\end{equation}
where $a$ denotes the ALP field, $f$ runs over Standard Model fermions, and $\Lambda_{f}$ parametrizes the strength of the ALP coupling to fermions. In the language of \cite{Eberhart:2025lyu}, this corresponds to the pseudoscalar coupling.

Although many models predict additional couplings to hadrons or gluons~\cite{Kim:2008hd}, it is common practice to consider each interaction channel independently in order to clearly assess experimental sensitivities. In this work, we assume that the dominant portal between the ALP and the Standard Model is provided by its couplings to charged leptons. Under this assumption, both the production and decay of the ALP at LDMX are primarily mediated by fermionic interactions. 

As discussed above, we are primarily interested in the experimental blind spots in the MeV–GeV mass range. This region contains a number of intriguing new-physics scenarios. Notable examples include the X17 anomaly~\cite{PhysRevLett.116.042501,Alves:2023ree}, as well as explanations of the long-standing $(g-2)$ discrepancy~\cite{Muong-2:2006rrc}. In addition, a number of arguments in favor of self-interacting dark matter point toward new physics in this mass range. Several realizations of the QCD axion and axion-like models with masses in this regime have also been actively explored in recent years (see~\cite{Murayama:2026ioh} as an example).

\subsection{Dark Photon Models}\label{sec_DP}

Dark photons, often denoted $A'$, are vector bosons associated with a new $U(1)_D$ gauge symmetry in the dark sector. In many well-motivated scenarios~\cite{Batell:2022dpx, Alexander:2016aln,battaglieri2017cosmicvisionsnewideas}, they interact with Standard Model fields primarily via the kinetic mixing portal:
\begin{equation}
\label{eq:ap_lagrange}
\mathcal{L} \;\supset\; -\frac{\varepsilon}{2} F'_{\mu\nu} F^{\mu\nu} \, ,
\end{equation}
where $F'_{\mu\nu}$ and $F_{\mu\nu}$ are the field strength tensors of the dark photon and the Standard Model photon, respectively, and $\varepsilon$ parametrizes the strength of the kinetic mixing. After diagonalizing the kinetic terms and rotating to the mass eigenbasis, the dark photon acquires a suppressed coupling to electrically charged fermions:
\begin{equation}
\label{eq:ap_interaction_lagrange}
\mathcal{L}_{\rm int} \;=\; - \varepsilon e \, \bar{f} \gamma^\mu f \, A'_\mu \, ,
\end{equation}
where $e$ is the electromagnetic coupling constant. This interaction allows the dark photon to be both produced and to decay into Standard Model particles, a mechanism that is utilized by many dark matter models and searches in the sub-GeV mass range (for example~\cite{Alexander:2016aln, Andreas:2012mt,NA482:2015wmo,LHCb:2017trq,BaBar:2014zli}).  

\subsection{The Light Dark Matter eXperiment (LDMX)}\label{sec_ldmx}
LDMX is designed to search for dark matter particles that interact very weakly with ordinary matter. The experiment uses an electron beam in the \SIrange[range-phrase=--]{4}{8}{GeV} range; for this study, we consider an \qty{8}{GeV} beam colliding with a thin tungsten target.
The detector concept allows each individual electron to be tagged and tracked as it passes through a thin target, closely spaced tracking planes, and a high-granularity silicon-tungsten calorimeter.
This setup enables precise reconstruction of both signal and background tracks, allowing a rigorous search for missing momentum and energy, which is a model-independent signature of dark matter production.

Beyond many other physics measurements~\cite{Berlin:2018bsc, akesson2025ldmxlightdark}, LDMX might also provide a unique opportunity to study very short-lived particle decays, with lifetimes on the order of $c\tau \sim \qty{10}{\micro \meter}$, in the clean environment of electron fixed-target collisions, where potential background is relatively straightforward to mitigate.
To fully realize LDMX's potential in these searches, some updates to the currently proposed setup (see~\cite{akesson2025ldmxlightdark}) might need to be included, as discussed later in Sec.~\ref{sec_RES}.

Similar experiments and searches for new physics are performed by HPS~\cite{PhysRevD.108.012015,PhysRevD.98.091101, 2k3q-9mhj}, NA64 at CERN~\cite{NA64:2017vtt}, and LHCb~\cite{Ilten:2015hya, Ilten:2016tkc}, among others. 
Throughout our analysis, we will borrow lessons from those efforts.

\section{Calculation} \label{sec_CALC}
The following discussion focuses exclusively on the axion case. However, the same considerations apply to the dark photon, with the replacement of the axion coupling scale $1/\Lambda$ by the parameter $\varepsilon$ (carefully keeping track of $m_f$ and $e$), as defined in Eq. \ref{eq:ap_interaction_lagrange}. The production cross section and decay length follow different expressions in the two scenarios, but aside from these differences, the search strategy outlined below remains unchanged.

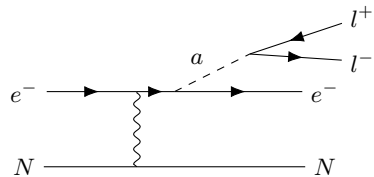
\begin{figure}[htb]
\centering

\hspace{1.2cm}

\begin{tikzpicture}
\begin{feynman}

\vertex (e_in)  at (-2, -1) {$e^-$};
\vertex (e_v1)   at (-0.5,  -1);
\vertex (e_v2)   at (0,  -1);
\vertex (e_out) at (2,  -1) {$e^-$};

\vertex (N_in)  at (-2, -2) {$N$};
\vertex (N_v)   at (-0.5, -2);
\vertex (N_out) at (2, -2) {$N$};

\vertex (p_ee_v)   at (1, -0.5);

\vertex (ep)     at (2.5, 0) {$l^+$};
\vertex (em)     at (2.5, -0.6) {$l^-$};

\diagram*{
(e_in) -- [fermion] (e_v1) -- [fermion] (e_v2) -- [fermion] (e_out),

(N_in) -- [solid] (N_v) -- [solid] (N_out),

(e_v1) -- [photon] (N_v),

(e_v2) -- [scalar, edge label=$a$] (p_ee_v),

(p_ee_v) -- [fermion] (em),
(p_ee_v) -- [anti fermion] (ep),
};

\end{feynman}
\end{tikzpicture}

\caption{Radiative axion production. The diagram also applies to the production of dark photons, replacing $a$ with $A'$. We also consider the additional diagram in which the axion is emitted before the photon exchange, corresponding to initial-state bremsstrahlung.}
\label{fig:axion_production}
\end{figure}

At LDMX, a highly collimated electron beam traverses the tracker. Owing to the nature of the beam, each electron can be studied independently, effectively resulting in a one-electron-at-a-time interaction with the target. The electron then interacts with a tungsten nucleus $N$ in the target and, for the signal process considered here, an axion is radiated from the incoming electron. The axion subsequently decays into a fermion pair, as illustrated in Sec. \ref{fig:axion_production}.

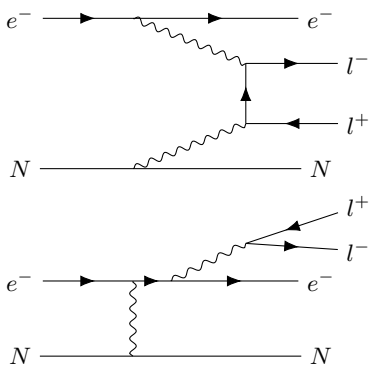
\begin{figure}[htb]
\centering
\begin{tikzpicture}
\begin{feynman}
\vertex (e_in)  at (-2, 0) {$e^-$};
\vertex (e_v)   at (-0.5, 0);
\vertex (e_out) at (2, 0) {$e^-$};

\vertex (N_in)  at (-2, -2) {$N$};
\vertex (N_v)   at (-0.5, -2);
\vertex (N_out) at (2, -2) {$N$};

\vertex (p_ep) at (1, -1.4);
\vertex (p_em) at (1, -0.6);

\vertex (ep)     at (2.5, -1.4) {$l^+$};
\vertex (em)     at (2.5, -0.6) {$l^-$};

\diagram*{
(e_in) -- [fermion] (e_v) -- [fermion] (e_out),
(N_in) -- [solid] (N_v) -- [solid] (N_out),

(e_v) -- [photon] (p_em),
(N_v) -- [photon] (p_ep),

(p_em) -- [anti fermion] (p_ep),

(p_em) -- [fermion] (em),
(p_ep) -- [anti fermion] (ep),
};

\end{feynman}
\end{tikzpicture}
\hspace{1.2cm}
\begin{tikzpicture}
\begin{feynman}

\vertex (e_in)  at (-2, -1) {$e^-$};
\vertex (e_v1)   at (-0.5,  -1);
\vertex (e_v2)   at (0,  -1);
\vertex (e_out) at (2,  -1) {$e^-$};

\vertex (N_in)  at (-2, -2) {$N$};
\vertex (N_v)   at (-0.5, -2);
\vertex (N_out) at (2, -2) {$N$};

\vertex (p_ee_v)   at (1, -0.5);

\vertex (ep)     at (2.5, 0) {$l^+$};
\vertex (em)     at (2.5, -0.6) {$l^-$};

\diagram*{
(e_in) -- [fermion] (e_v1) -- [fermion] (e_v2) -- [fermion] (e_out),

(N_in) -- [solid] (N_v) -- [solid] (N_out),

(e_v1) -- [photon] (N_v),

(e_v2) -- [photon] (p_ee_v),

(p_ee_v) -- [fermion] (em),
(p_ee_v) -- [anti fermion] (ep),
};

\end{feynman}
\end{tikzpicture}

\caption{Top: Bethe--Heitler trident. Bottom: Radiative trident. In both cases, we also consider all possible diagram topologies.}
\label{fig:rad_bh_tridents}
\end{figure}

The background for such processes consists primarily of two contributions: radiative trident diagrams and Bethe–Heitler processes (see Fig. \ref{fig:rad_bh_tridents}), with the latter dominating the production rate.
Other sources of background, such as production through off-shell Higgs or $Z$ bosons, may also contribute; however, their cross sections are strongly suppressed by the fourth power of the mediator masses. For this reason, these contributions are neglected in the present analysis.

For both signal and background events, the cross sections and invariant-mass distributions were generated using \texttt{MadGraph5}~\cite{Alwall_2014}. 
The \texttt{MadGraph5} $A'$ model was originally developed for HPS on the basis of \cite{Morrissey_2009,PhysRevD.80.035008,PhysRevD.80.015003,Bjorken_2009}.
The model has been adapted for this work, and an axion generator has been developed on the same basis.
The signal samples are generated at different mass points in the range of \qtyrange{25}{500}{MeV}. For each mass, the cross section is determined over \num{10000} generated events. 
This step is simplified by simulating only the production of an axion or dark photon from the incoming electron beam, excluding the subsequent decay, at a fixed coupling strength ($1/\Lambda$, $\varepsilon$).
The cross sections and lifetimes of the signal particles are reweighted appropriately at a later stage of the analysis. 

In general, the signal cross section exhibits a clear dependence on the coupling squared, \(\sigma_s \sim 1/\Lambda^2\) for the axion or \(\sigma_s \sim \varepsilon^2\) for the dark photon. However, due to the nuclear form factors associated with interactions between the tungsten nucleus and electrons, a non-trivial mass dependence arises. This dependence is ultimately extracted directly from the simulated data. 

For the background processes, $e^+e^-$ and $\mu^+\mu^-$ pairs, originating from radiative and Bethe-Heitler processes as shown in Fig. \ref{fig:rad_bh_tridents}, are generated.
In a first pass, the $l^+l^-$ pairs are generated with minimal requirements on the kinematics of the outgoing particles. A small section of the phase space must be excluded because of a divergence in the Bethe-Heitler cross section (see, e.g.,~\cite{Bjorken_2009}) to ensure predictable behavior of \texttt{MadGraph5}.
Additionally, for each of the two background processes, two extra samples are produced, requiring $m_\text{inv}^{l^+l^-} > \qty{0.01}{GeV}$ and $m_\text{inv}^{l^+l^-} > \qty{0.05}{GeV}$, respectively, to increase the coverage of the high-invariant-mass tails. 
All discussed cuts are summarized in Tab. \ref{tab:MG5_bkg_cuts}.

\begin{table}[ht]
    \centering
    \begin{tabular}{cc}
    \toprule
    \texttt{MadGraph5} constraint & purpose \\
    \midrule
    \makecell{
      $E_{l^+}+E_{l^-} > \qty{0.2}{\giga\electronvolt}$ \\[1mm]
      $\dfrac{\min(|\theta_{l^+}|, |\theta_{l^-}|)}{\max(|\theta_{l^+}|, |\theta_{l^-}|)} > 0.001$
    } 
    & \makecell{Controlling Bethe-Heitler\\ cross section} \\
    $m_\text{inv}^{l^+l^-} > 0.01~(0.05)~\mathrm{GeV}$
    & \makecell{Increasing statistics \\ in high $m_\text{inv}$ tail} \\
    \bottomrule
    \end{tabular}
    \caption{MadGraph5 constraints and their purposes}
    \label{tab:MG5_bkg_cuts}
\end{table}

Once the production cross sections for both signal and background are obtained as functions of the coupling and the axion mass, we can then study strategies that combine resonance reconstruction (in invariant mass) and vertexing to mitigate background. Vertexing, the identification of displaced decay vertices, is particularly effective for constraining processes with moderately long-lived particles, allowing sensitivity to smaller couplings.
Resonance searches, on the other hand, exploit large event yields to constrain larger couplings, and are most effective for prompt processes that cannot be distinguished from the background solely by vertex position. Both methods are implemented simultaneously in our study, and their details are discussed below.

\begin{figure*}[ht]
    \centering
    \begin{minipage}{0.48\linewidth}
        \centering
        \includegraphics[width=\linewidth]{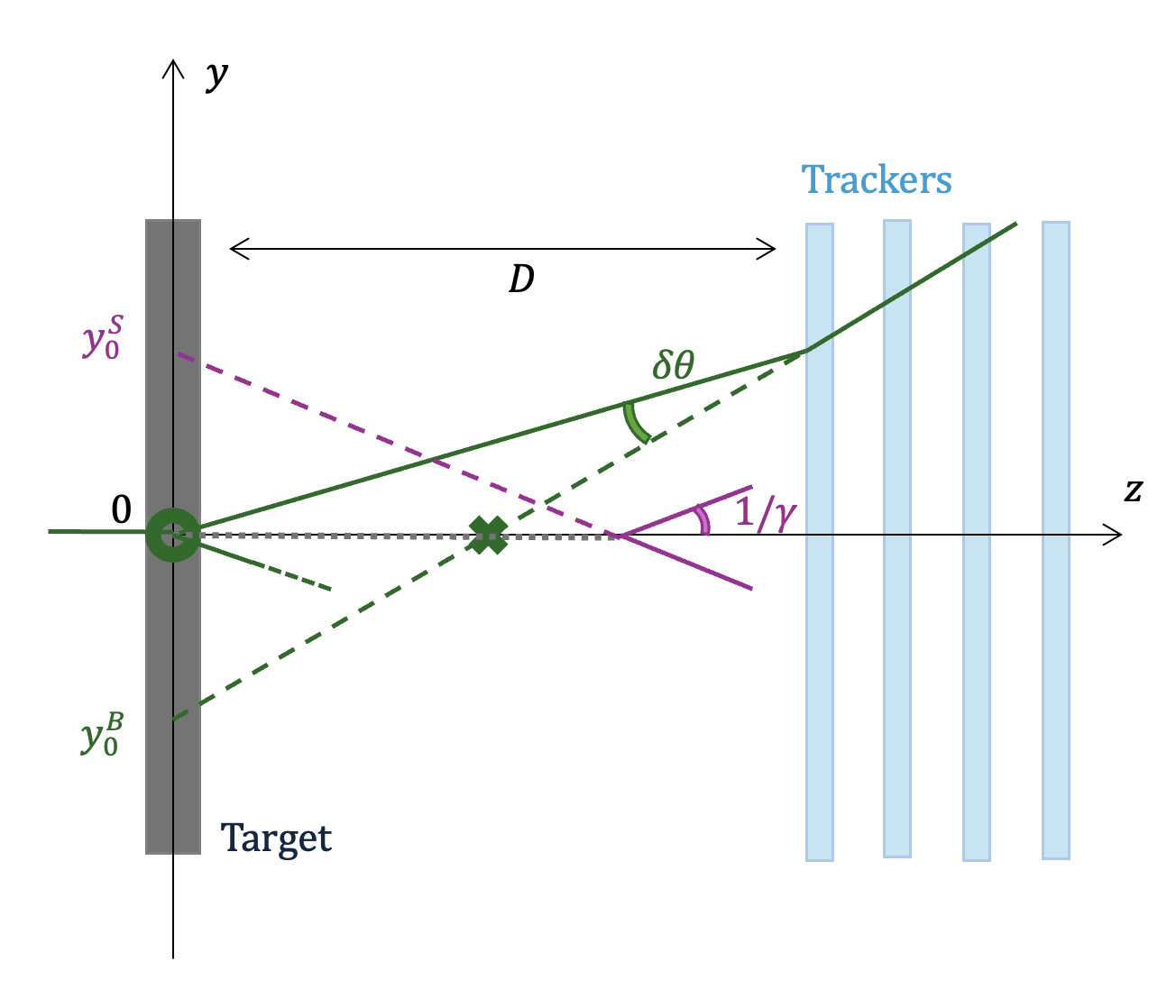}
    \end{minipage}\hfill
    \begin{minipage}{0.48\linewidth}
        \centering
        \includegraphics[width=\linewidth]{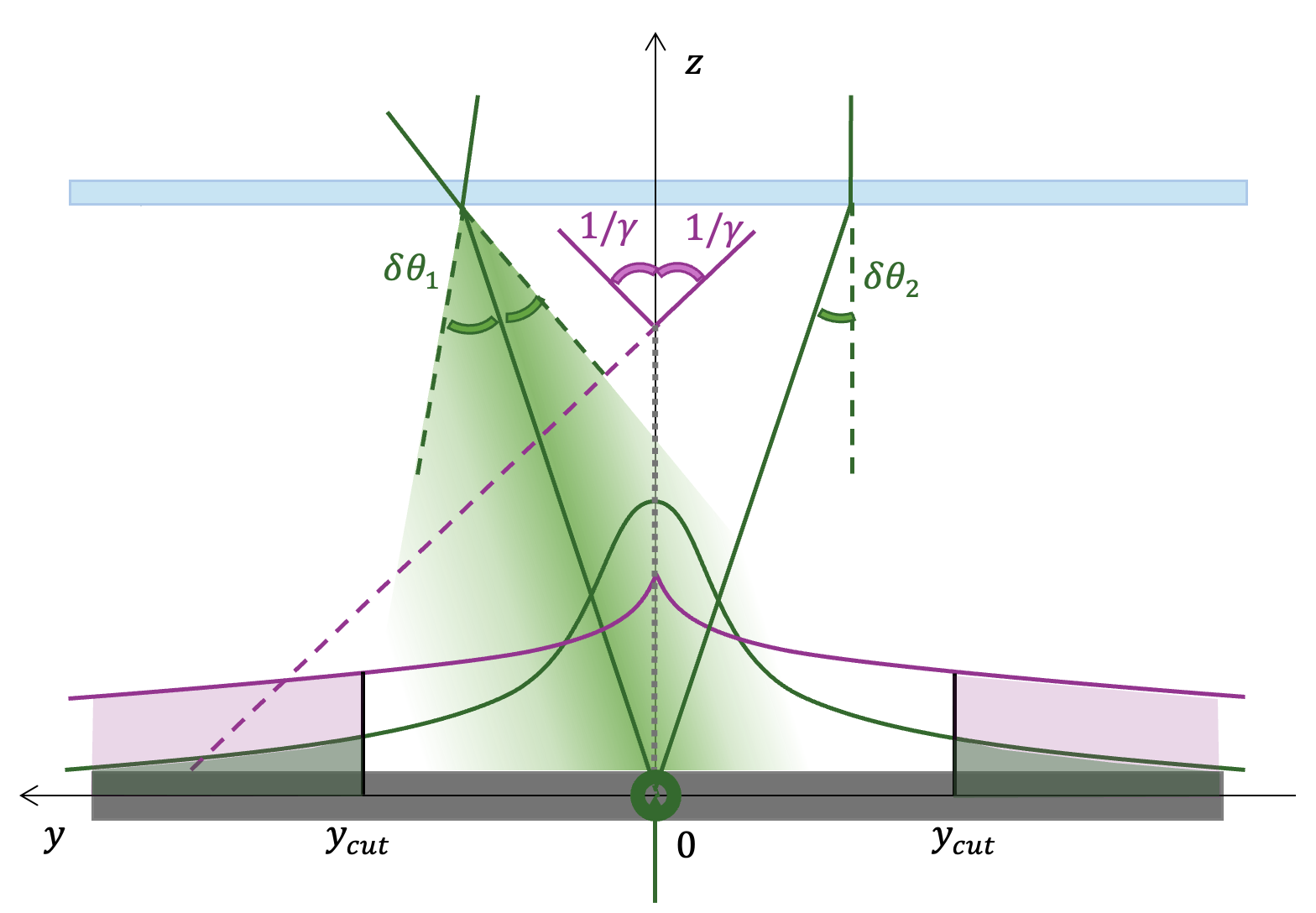}
    \end{minipage}
    \caption[Schematic layout of an LDMX-like experiment]{
    Schematic layout of an LDMX-like experiment. The electron beam collides with the target (gray, made of tungsten with a thickness corresponding to $\qty{5}{\%}$ of a radiation length) at $y = z = 0$. The distance between the target and the first tracking layer (light blue, $\sim \qty{100}{\micro m}$ in thickness) is denoted by $D$ (which is taken to be \qty{1}{cm}). Background fermions and axions are produced at the beamspot, and their trajectories are shown by the green lines and the gray dotted line, respectively. The fermions produced in axion decays are shown in purple. For simplicity, the electron coming from the beam is not shown after the collision.
    In the left panel, the dashed lines represent the reconstructed tracks. Due to multiple scattering, the apparent production point of the background electrons is displaced, as indicated by the green cross. The quantities $y_0^S$ and $y_0^B$ denote the impact parameters of the reconstructed signal and background tracks, respectively.
    The right panel shows the same geometry rotated by $90^\circ$ and illustrates the resulting distributions in $y_0$ for signal and background electrons. For the background (green), fermions produced at $y = z = 0$ can independently scatter with angles $\delta\theta$ in the trackers, drawn from the green probability density function centered at zero. The shaded cone around $\delta\theta_1$ illustrates the stochastic nature of this scattering. Since $\delta\theta_1$ and $\delta\theta_2$ are independent, the probability for both electrons to satisfy a given $y_0$ cut is given by the square of the corresponding PDF integral (as in Eq.~\eqref{eq:B_fin}).
    For the signal (purple), the electron angles are, to first approximation, determined by the boost factor of the axion and are therefore correlated. As a result, the integrated PDF for $y_0^S$ enters linearly, as shown in Eq.~\eqref{eq:S_fin}. The shapes of both background and signal PDFs are discussed in Secs.~\ref{sec:calc:bkg_model} and~\ref{sec:calc:signal_model}. We neglect the effect of Molière scattering on the signal electrons; this is a conservative choice, since the signal PDF is monotonically decreasing in $|y_0|$ and any additional smearing would only increase the signal yield.
    In this schematic, opening angles, vertical and horizontal distances have been exaggerated for clarity. For small angles, the approximations
    $y_0^S \simeq \frac{1}{\gamma}\, \gamma c\tau$
    and
    $y_0^B \simeq \delta\theta \, D$
    are valid. }
    \label{fig:scattering_schematic}
\end{figure*}

\subsection{Background model}\label{sec:calc:bkg_model}

We estimate signal sensitivity using a background model guided by the experience of the HPS experiment. Rather than performing a full simulation or 3D event reconstruction, we conservatively estimate background rejection using vertex identification, specifically the transverse impact parameter, obtained by projecting tracks onto the plane transverse to the beam (which corresponds to $y_0$ in Fig \ref{fig:scattering_schematic}). A more detailed study by LDMX will be needed to improve upon our estimates, but the results presented here provide a useful guide for motivating the required detector performance and for scoping more computationally expensive full simulation studies.

\subsubsection{\label{sec:calc:bkg_model:vtx} Vertexing}

Thanks to the tracker layers located very close to the LDMX target, it is possible to reconstruct the trajectories of the outgoing fermions and identify the location where the fermion pair was produced.
We conventionally define the $z$ axis as parallel to the beam momentum and centered at the target position, and $y$ perpendicular to $z$ going through the target plane parallel to the central magnetic field (see the left part of Fig. \ref{fig:scattering_schematic}).
From the nature of Bethe--Heitler and radiative processes, most of the reconstructed vertices are expected to be located at $y=z=0$, corresponding to fermion pairs produced directly in the target.

In practice, however, two experimental effects can shift the reconstructed value of the impact parameter $y_0$ for background events: multiple scattering of the electrons in the tracking material and the finite impact parameter resolution of the detector. For this reason, a reliable background model to describe these effects must be defined. Following the convention shown in Fig. \ref{fig:scattering_schematic}, we define $y_0^B$ as the distance from the origin to the intersection point of the reconstructed background fermion track with the plane of the target. The distribution of $y_0^B$ is primarily modeled by multiple scattering and Rutherford scattering effects.

\begin{figure*}[t]
    \begin{minipage}{0.48\linewidth}
        \centering
        \includegraphics[width=\linewidth]{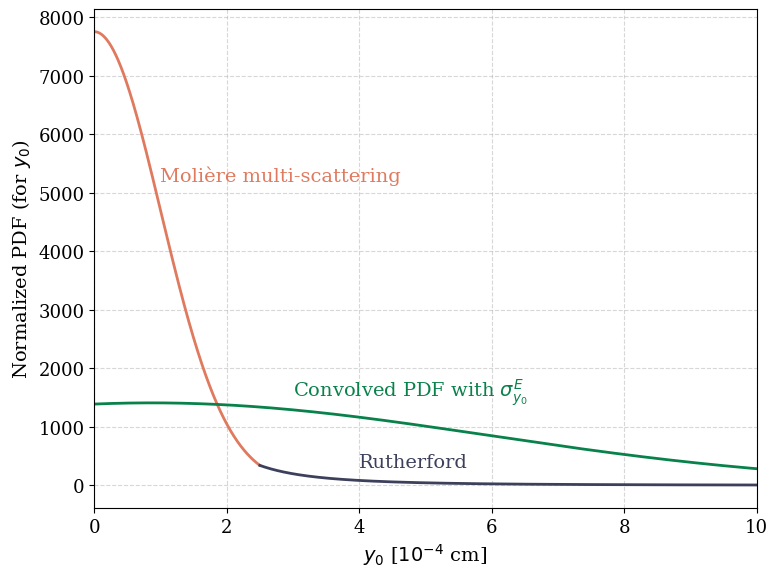}
    \end{minipage}\hfill
    \begin{minipage}{0.48\linewidth}
        \centering
        \includegraphics[width=\linewidth]{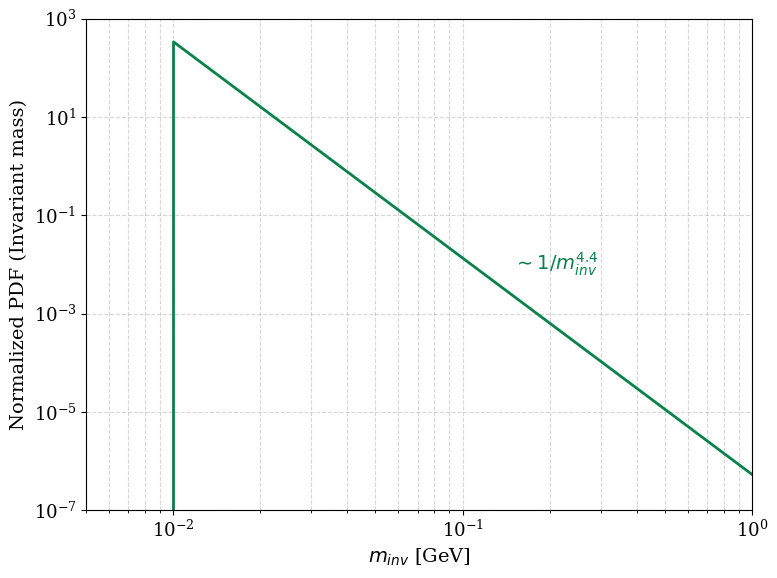}
    \end{minipage}
    \caption{Modeled background distributions in $y_0^B$ (left) and invariant mass (right). For the $y_0^B$ distribution, the orange curve (extending up to $y_0^B = \qty{2.5}{\micro\meter}$) describes the region of the background where the impact parameter $y_0$ is dominated by a Gaussian Molière multiple-scattering component, while the blue curve represents the power-law tail arising from Rutherford scattering. The transition value $y_0 = \qty{2.5}{\micro\meter}$, corresponding to approximately $2.5$ standard deviations of the Gaussian core, is consistent with \cite{Bethe:1953va}. This choice should be regarded as a working assumption (compatible with the experience of experiments like HPS), but we note that the detector configuration and data-taking conditions can affect the location of the transition from Gaussian to power law. For the invariant mass distribution, we simply fit the simulated data with a power-law function, applying a lower cut at \qty{10}{MeV}, and find good agreement with a distribution proportional to $m_{\mathrm{inv}}^{-4.4}$.}
    \label{fig:background_pdfs}
\end{figure*}

The basic idea is that a fermion, while propagating from the target to the first tracking layer, can undergo (multiple) scatterings with silicon atoms, resulting in a small change of direction by an angle $\delta\theta$.
This effect is less significant for subsequent layers as they are much closer to one another and can better account for corrections.
The scattering of charged fermions traversing matter at small angles is well described in Sec.~34 of \cite{ParticleDataGroup:2024cfk}. 
We model our background such that for small values of $\delta\theta$, the angular distribution is approximately Gaussian and centered at zero, while for larger angles it follows a power-law tail dictated by Rutherford scattering; see also Fig. \ref{fig:background_pdfs}. 

As illustrated in Fig. \ref{fig:scattering_schematic}, the resulting value of $y_0^B$ for a given track is given by $y_0^B \simeq \delta\theta \, D$, where $D$ denotes the distance between the target and the first tracking layer. We use this model as our $y_0^B$ PDF.

Clearly, larger values of $y_0$ correspond to larger reconstructed values of $z_{\rm vtx}$ (the apparent location of the pair production point along the $z$-axis), which can cause a background fermion pair to be misidentified as originating from a displaced decay rather than being produced directly in the target: a misidentification that mimics the signal.
Our vertexing requirement, therefore, consists of demanding that both fermions in a background event have values of $y_0^B$ larger than a given threshold $y_{\rm cut}$, commensurate with what is expected from a typical signal event.

Since each fermion can scatter independently, the probability for a background event to satisfy this requirement is given by the square of the single-fermion probability for $y_0$ to exceed $y_{\rm cut}$.
This single-fermion probability is obtained by integrating the modeled background PDF from $y_{\rm cut}$ to infinity.
Fig. \ref{fig:scattering_schematic} illustrates this effect, where the distributions of $\delta\theta_1$ and $\delta\theta_2$, which directly map onto the $y_0$ distributions, are independent and centered at zero.

The remaining ingredient required to model the experimental $y_0$ distribution for the background is the detector resolution on $y_0$. Operationally, this is implemented by convolving the theoretical background distribution shown in orange and blue in Fig. \ref{fig:background_pdfs} with a Gaussian of width $\sigma_{y_0}^E$. The resulting distribution is then compared to the $y_0^B$ distribution of the signal in order to define a vertex cut (see Sec.~\ref{sec_sig}). This cut, when combined with an invariant mass selection discussed in the following sections, is optimized to maximize the signal significance. 

This convoluted PDF, which we denote as $P_B^V$, is shown in green in the left panel of Fig. \ref{fig:background_pdfs} and, in an exaggerated schematic form, also in green in Fig. \ref{fig:scattering_schematic}. 
We also emphasize that, as long as the electron pair is produced at $y = z = 0$ (see Sec.~\ref{sec:calc:bkg_model:additional_BG} for the alternative case), any opening angle of the electron pair, $\alpha$, does not affect the impact parameter $y_0^B$ at leading order.

\subsubsection{\label{sec:calc:bkg_model:resonance}Resonance Search}

A resonance search in the invariant mass spectrum can significantly constrain regions of parameter space corresponding to prompt decays, which cannot be efficiently probed using vertexing cuts alone. For the background, the invariant mass distribution is modeled directly from simulated events. After applying a lower invariant mass cut of \qty{10}{MeV}, the background spectrum is well described by a power-law distribution shown in Fig. \ref{fig:background_pdfs}.

\subsubsection{\label{sec:calc:bkg_model:additional_BG}Additional Backgrounds}

We first consider possible corrections to the impact parameter $y_0^B$ arising from situations in which the electron pair exits the target with a non-zero transverse displacement, i.e., events for which at $z = 0$ we have $y \neq 0$. Several background processes could, in principle, lead to such configurations. However, we argue that their contribution is negligible, especially once events whose production vertex is reconstructed within the target are discarded before applying any vertex cut (see Sec.~\ref{sec_geometry} for details).

First, if the pair is produced at the upstream edge of the target with an opening angle $\alpha$, it may exit the target with $y \neq 0$. For a target thickness corresponding to $\qty{5}{\%}$ of a radiation length, this effect is small: the induced shift $\delta y_0^B$ is smaller than the impact parameter resolution $\sigma_{y_0}^E$. While this estimate depends in principle on the energy and invariant mass (through $\alpha$), it remains valid across the region of interest considered here ($\lesssim \qty{500}{MeV}$), especially if full three-dimensional reconstruction is taken into account.

Additional contributions from multiple scattering inside the target are even more subdominant. Moreover, such events would still be rejected by requiring that the reconstructed production vertex lies outside the target thickness.

Another potential background arises from rare configurations in which one background electron is produced with a large opening angle $\alpha$ and subsequently undergoes Rutherford scattering inside the target by an angle of order $2\alpha$, effectively crossing the trajectory of the second electron outside the target. Although such configurations are kinematically possible, they are highly suppressed, particularly for a thin target, and can therefore be safely neglected.

As explained at the end of Sec.~\ref{sec_geometry}, these events can be further reduced by imposing an additional geometric constraint: the reconstructed impact-parameter points of the two outgoing electrons and the incoming beam electron should be approximately collinear in the transverse plane (within detector resolution and accounting for multiple scattering in the target), as expected from momentum conservation.

Finally, additional backgrounds may arise from electrons originating from muon or pion decays. We do not consider these processes explicitly, as we expect them to be efficiently suppressed through beam-energy, missing-energy, and kinematic selections.

\subsection{Signal Model}
\label{sec:calc:signal_model}

The signal model is considerably simpler than the background model described above. The variable $y_0^S$ follows a simple exponential distribution since,  based on Fig. \ref{fig:scattering_schematic},  $y_0^S$ is effectively equivalent to the unboosted decay length, $c\tau$. This can be seen from the relation $y_0^S = D_a\,\theta^S$, where $D_a$ is the distance traveled by the axion before decaying, $D_a = \gamma c\tau$, and $\theta^S$ is the typical opening angle of the decay products, which for a boosted decay is approximately $1/\gamma$. As a result, the boost factors cancel, and the reconstructed displacement $y_0^S$ directly traces the intrinsic decay length of the particle.

In theory, the $y_0^S$ distribution for the signal is therefore described by an exponential decay, as illustrated by the purple curve in Fig. \ref{fig:scattering_schematic}, which we denote as $P_S^V$. In practice, however, two effects can modify the shape of this distribution: multiple scattering and the finite experimental resolution. These are the same effects that affect the background electrons.

Both effects act as a smearing of the true distribution. In particular, they can make some low-$y_0^S$ events appear more displaced, while shifting some high-$y_0^S$ events closer to the origin. Since $P_S^V$ is monotonically decreasing in $|y_0|$, with many more events at small $y_0$ than at large $y_0$, including this smearing would increase the number of signal events passing a $y_{\rm cut}$ requirement. For this reason, and in order to remain conservative, we choose not to include the effects of multiple scattering and experimental resolution in the signal $y_0^S$ model. Our conservative strategy is to maximize the expected background yield while minimizing the signal yield. Because smearing would increase the event acceptance, we apply it only to the background and not to the signal.

The invariant mass distribution is even simpler: since the axion is produced on-shell, it is well described by a Gaussian centered at the axion mass, with a width dominated by the experimental mass resolution $\sigma_m^E$. There is, however, one correction to this simple model, which concerns axions that decay extremely promptly and produce the fermion pair within the thickness of the detector material. In this case, additional multiple-scattering effects arise, similar to those discussed in Sec. \ref{sec:calc:bkg_model:vtx}. This correction is taken into account by geometrically adding the experimental mass resolution $\sigma_m^E$ to the momentum uncertainty induced by this effect, resulting in an effective invariant-mass resolution $\sigma_m$ used in the signal model.

\subsection{Geometric Constraints} \label{sec_geometry}

There are additional geometric constraints that must be taken into account. In particular, we require that, for signal events, the electron pair production vertex is reconstructed within the tracking volume, i.e., well before the end of the tracker located at $L_{\rm tracker} = \qty{20}{cm}$. 
This requirement effectively multiplies the signal yield by the integrated decay probability up to, to be conservative, $D_T = \frac{1}{2}L_{\rm tracker} = \qty{10}{cm}$,
\begin{equation}
f_G^{0D} = \int_0^{D_T} \frac{e^{-z/\gamma c \tau}}{\gamma c \tau} \, dz \, .
\end{equation}

When applying the vertex cut, in order to suppress background events that can appear with a large impact parameter $y_0$ already at $z=0$ (for instance due to a large production angle $\alpha$ or hard Rutherford scattering inside the target), we further require that the electron pair is produced outside the target, i.e., at $z > d_t$, where $d_t$ is the target thickness. This introduces a new factor for the signal,
\begin{equation}
f_G^{dD} = \int_{d_t}^{D_T} \frac{e^{-z/\gamma c \tau}}{\gamma c \tau} \, dz \, .
\end{equation}

This latter requirement significantly reduces the sensitivity of a pure resonance search. For this reason, our analysis adopts a hybrid strategy: we use $f_G^{0D}$ when the characteristic decay length $\gamma c\tau$ is smaller than the target thickness $d_t$, and $f_G^{dD}$ when $\gamma c\tau$ exceeds $d_t$. In the regime $\gamma c\tau < d_t$, we do not apply any vertexing cut and rely solely on the resonance search.

This treatment is intentionally conservative. We apply these geometric suppression factors only to the signal and not to the background. In practice, this corresponds to assuming that events produced inside the target, in the context of a vertexing analysis, would likely be discarded in a realistic experimental setup. These events, however, remain highly relevant in a purely resonant search.

A different effect that deserves comment is the role of the random production angle $\alpha_a$ of the two signal electrons in the rest frame of the axion, and its impact on the boosted opening angle. In this work, we approximate the opening angle of the boosted electrons from the axion decay as $\sim 1/\gamma$. However, in the rest frame, the electrons are produced back-to-back with a uniform angular distribution in $\alpha_a$. When boosting to the lab frame, this introduces a correction to the opening angle which, in the highly boosted regime, scales approximately with $\tan(\alpha_a/2)$. Averaging over many events, this effect can be effectively absorbed into the definition of the typical opening angle. We have explicitly tested different scenarios and find that, as long as we restrict ourselves to the fiducial and approximate interval $0.05\pi < \alpha_a < 0.95\pi$, corresponding to an effective $\sim \qty{10}{\%}$ signal loss, which we incorporate as $f_G^{\alpha_a} = 0.9$, our sensitivity reach is not significantly affected.

Lastly, we emphasize that our analysis is effectively two-dimensional, while in a realistic setup, one could further improve the background rejection by implementing a fully three-dimensional selection based on track angles in the transverse plane. In particular, one could define an angle $\varphi$ in the $x$--$y$ plane at $z=0$, constructed from the points $T_1$ and $T_2$, which are the intersections of the two electron tracks with this plane, and the point $z_0$ where the incoming beam electron crosses it. In this picture, one could apply both a radial cut $r_{\rm cut}$ (the two-dimensional analogue of $y_{\rm cut}$) and an angular cut $\varphi_{\rm cut}$. The latter would require that the two outgoing electron tracks and the incoming beam track lie approximately in the same transverse plane, which corresponds to the second electron being emitted at an angle $\varphi \simeq \pi$ with respect to the first, with the beam direction defining the axis of rotation. In practice, this condition would be satisfied within some tolerance $\delta\varphi$, determined by experimental resolution and multiple scattering in the target and tracking layers. Such a cut would provide additional background suppression. However, we do not implement this analysis here, as it would require a full three-dimensional Monte Carlo simulation to accurately model the angular distributions at this level.

\subsection{Significance} \label{sec_sig}

The goal of this work is to demonstrate that LDMX can achieve significant sensitivity to axions (and dark photons) over a wide region of parameter space. To this end, we implement the signal and background models described above and apply selection cuts both on $y_0$ (vertexing) and on the invariant mass.

For the vertexing selection, we only accept values of $y_0$ corresponding to a sufficiently displaced decay position $z_{\rm vtx}$. Operationally, this is achieved by integrating the signal and background probability density functions from a minimum value $y_{\rm cut}$ to $+\infty$. 

The resonance search selection consists of a window cut applied to the invariant mass distribution. The window is centered at the axion mass $m_a$ and has a half-width given by $f_m \sigma_m$, where $f_m$ is a free parameter chosen to maximize the significance. In the approximation of a flat background, the optimal choice is $f_m \simeq 1.4$.

Applying both selections simultaneously, the expected signal and background yields can be written as

\begin{widetext}
\begin{equation}
S(m_a, \Lambda, y_{\rm cut}, f_m)  =
\sigma_a(m_a,\Lambda)\,\mathcal{L}_{\rm int}
\int_{m_a-f_m\sigma_m}^{m_a+f_m\sigma_m} dm\, P_S^M(m) \, \, f_G^{\alpha_a}\times
\begin{cases}
\displaystyle
f_G^{0D}, &\gamma c\tau < d_t,\\[1.2ex]
\displaystyle
f_G^{dD}\int_{y_{\rm cut}}^{+\infty} dy_0\, P_S^V(y_0),
 &\gamma c\tau > d_t.
\end{cases}
\label{eq:S_fin}
\end{equation}

\begin{equation}
B(m_a, y_{\rm cut}, f_m)  =
\sigma_b\,\mathcal{L}_{\rm int}
\int_{m_a-f_m\sigma_m}^{m_a+f_m\sigma_m} dm\, P_B^M(m) \times
\begin{cases}
\displaystyle
1, & \gamma c\tau < d_t,\\[1.2ex]
\displaystyle
\left(\int_{y_{\rm cut}}^{+\infty} dy_0\, P_B^V(y_0)\right)^2,
& \gamma c\tau > d_t.
\end{cases}
\label{eq:B_fin}
\end{equation}
\end{widetext}

Here, $\sigma_a$ and $\sigma_b$ denote the signal and background production cross sections, respectively, $\mathcal{L}_{\rm int}$ is the integrated luminosity, and $P_{S,B}^V$ and $P_{S,B}^M$ represent the vertexing and invariant-mass probability density functions for signal and background. As a reminder, the vertex cut enters quadratically for the background but only linearly for the signal: in a background event, both fermions must independently satisfy the $y_0 > y_{\rm cut}$ requirement, resulting in a squared probability, whereas for the signal the two fermion tracks originate from the same displaced decay and are therefore correlated. As discussed in Sec.~\ref{sec_geometry}, the vertex cut is applied only when the characteristic boosted decay length exceeds the target thickness; the geometric suppression factors $f_G^{0D}$, $f_G^{dD}$, and $f_G^{\alpha_a}$ are applied to the signal throughout.
The overall significance is then quantified using the standard Asimov significance for exclusion as in \cite{Bhattiprolu:2020mwi}, and $y_{\rm cut}$ and $f_m$ are found to maximize this expression. We also require that at least one signal event pass this selection when evaluating the significance. 

\section{Results} \label{sec_RES}

\begin{figure*}[!t]
    \centering
    \begin{minipage}{0.48\linewidth}
        \centering
        \includegraphics[width=\linewidth]{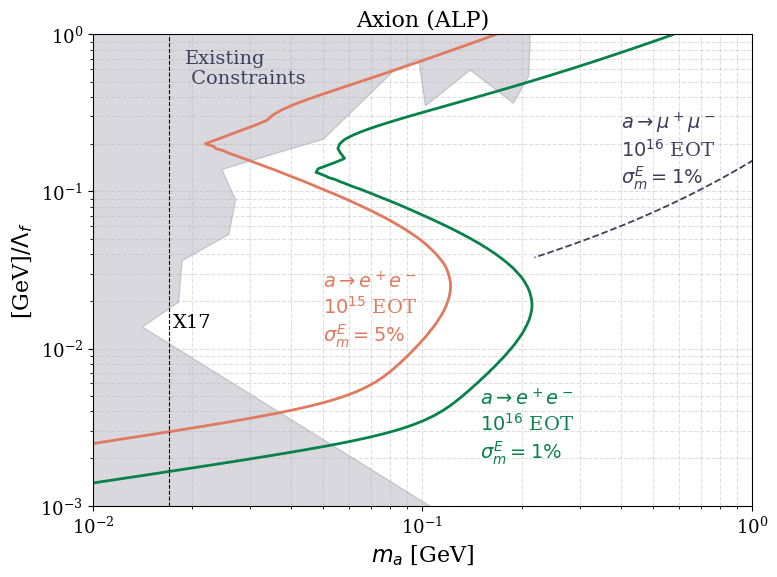}
    \end{minipage}\hfill
    \begin{minipage}{0.48\linewidth}
        \centering
        \includegraphics[width=\linewidth]{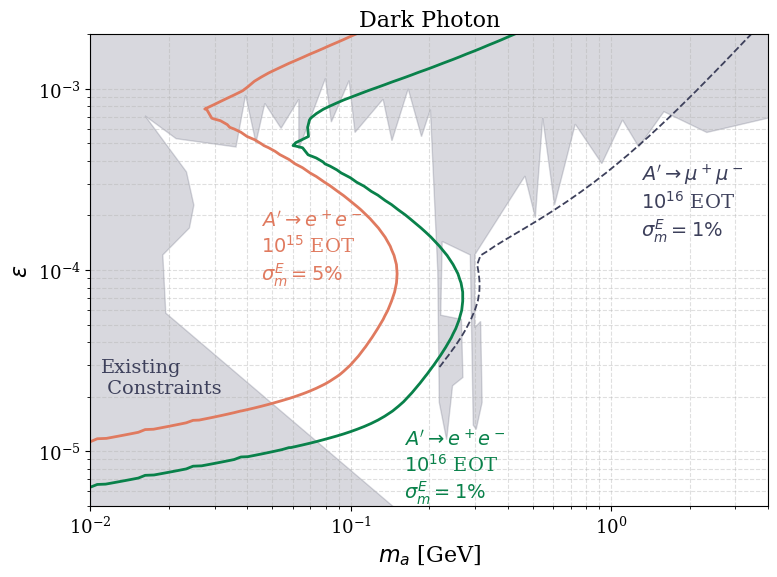}
    \end{minipage}
    \caption{In both plots (axion on the left and dark photon on the right), two experimental scenarios are considered, and the corresponding projected $\qty{95}{\%}$ exclusion regions are shown. The orange curves (leftmost solid lines in both plots) correspond to a pessimistic scenario, assuming $10^{15}$ electrons on target (EOT) and a mass resolution of $\qty{5}{\%}$. The more optimistic scenario is shown by the green curves, which assume $10^{16}$ EOT and a mass resolution of $\qty{1}{\%}$. Both of these are for $a/A' \rightarrow e^+ e^-$. The dashed blue curves indicate the projected exclusion for an axion or dark photon decaying predominantly into muons. The shaded gray regions represent existing constraints for an electron-philic axion or dark photon, as reported in \cite{Eberhart:2025lyu, Batell:2022dpx}. For the axion, these include the beam dump experiments E141, E137, and $\nu$Cal, the BaBar monophoton search, and LEP \cite{Belle-II:2020jti, Jaeckel:2015jla}; for the dark photon, constraints come from the E141, Orsay, and E137 beam dumps, NA48/2, LHCb, BaBar, and others \cite{Andreas:2012mt, NA482:2015wmo, LHCb:2017trq, BaBar:2014zli}.}
    \label{fig:predicted_reach}
\end{figure*}
Applying the cuts outlined in Sec.~\ref{sec_CALC}, our search strategy is sensitive (\qty{95}{\%} confidence level) to a whole new region in parameter space for both the axion and the dark photon. Our results, shown in Fig.~\ref{fig:predicted_reach}, cover a wide range in couplings in the sub-GeV range.
For the axion case, using these parameters ($d_t = \qty{5}{\%}$ of the radiation length, $D = \qty{1}{cm}$, $D_T = \qty{10}{cm}$, $\sigma^E_{y_0} = \qty{10}{\micro m}$), LDMX could cover the X17 excess region, extending to masses well above \qty{100}{MeV}. A similar conclusion holds for the dark photon scenario. Two distinct sensitivity regimes can be clearly identified for both the axion and the dark photon.
At smaller couplings, vertexing cuts dominate the sensitivity and are responsible for the deepest exclusion reach, around couplings of order $2 \times 10^{-2}$ for the axion (and $\sim 10^{-4}$ for the dark photon).
At larger couplings, invariant-mass resonance searches become the leading handle, allowing the experiment to probe the higher-coupling region of parameter space. 

The dashed blue muon line has a different interpretation in the axion and dark photon plots. In the axion case, it serves only as an indicative exclusion reach for decays into muons, since existing constraints are typically shown for the electron-philic scenario.
In contrast, for the dark photon, the structure of the coupling is the same for electrons and muons; as a result, the muon constraints are directly applicable and provide equally strong bounds on the kinetic-mixing parameter $\varepsilon$. 
In both plots, the muon line extends down to masses of approximately twice the muon mass.

As noted in Sec.~\ref{sec_ldmx}, LDMX's nominal setup may need adjustment to realize its full potential for visible searches; for our analysis, we adopt an LDMX-like configuration with a tungsten target thickness of \qty{5}{\%} of a radiation length.
A thinner target would lead to a more precise resonance search since particles are less likely to decay within the target, and the momentum reconstruction is cleaner.
Furthermore, scattering in the target material contributes to the broadening of the vertical impact parameter distributions, both for signal and backgrounds. 
Decreasing the target thickness to $\mathcal{O}(\qty{1}{\%})$ of a radiation length would reduce these effects.
However, this comes at the cost of reduced luminosity and hence significance, since the significance scales as $\sqrt{\mathcal{L}}$.
We also note that the LDMX trigger system is designed predominantly for missing momentum signatures and will not necessarily capture the visible decay events discussed in our analysis.
Given LDMX's flexible trigger architecture, which already accommodates visible decays in deeper detector layers~\cite{akesson2025ldmxlightdark}, appropriate trigger requirements for prompt and short-lived visible decays could be incorporated.

\section{Acknowledgments}
We thank Natalia Toro, Timothy Nelson and Daniele Spier Moreira Alves for their feedback and fruitful discussions.
We are grateful to Matthew Gignac and Emrys Peets for sharing their insights on the displaced vertex and resonance search analysis techniques.
The research reported here is supported by Stanford University under Contract No. DE-AC02-76SF00515 with the U.S. Department of Energy, Office of Science, Office of High Energy Physics; by the National Science Foundation under Grant No. PHY-2310429; by the Simons Foundation through Investigator Award No. 824870; and by the John Templeton Foundation under Award No. 63595.

\bibliographystyle{apsrev4-2}
\bibliography{bib}

\end{document}